\documentclass[12pt]{iopart}
\usepackage{bm}
\usepackage{graphicx}
\begin{document}
\title[The Brillouin Instability of intense laser in relativistic plasmas ]{The Brillouin Instability of intense laser in relativistic plasmas \footnote{Supported by Ph.d Foundation
of China Education (Grant No. 20020027006)}}

\author{Hong-Yu Wang$^{1,2}$
 and Zu-Qia Huang$^1$ }

\address{$^1$Beijing Normal
University, Institute of Low Energy Nuclear Physics, Beijing,
100875, China}
\address{$^2$Anshan Normal University, Department of Physics, Anshan,
114005, China}

\begin{abstract}
This paper studies the propagation of intense laser in plasmas in
weak relativistic region($0.1<a_0^2<0.5$) using the quasi-periodic
approximation. Effective Lorentz factor and density wave effects
are calculated in detail. The relativistic correction on
stimulated Brillouin instability is investigated in the rest
parts. The coupled dispersion relations of Stimulated Brillouin
Scattering(SBS) are obtained and investigated numerically.
\end{abstract}


\section{Introduction}

The propagation of intense laser through plasmas is an important
concern in the laser-driven inertial confinement fusion, the
laser-plasma accelerators, the x-ray laser and other physical
problems \cite{1}-\cite{5}. In General, the amplitude of laser can
be described by the normalized vector potential $\{\bf{a},\phi\}$,
where $\mathbf{a}=\frac {e\mathbf{A}}{m_0c^2}, \phi=\frac
{e\Phi}{m_0c^2}$. While $a\ll1$, the propagating equation of laser
is linear. But many non-linear effects will appear when $a$
increases , such as density wave effects, relativistic mass
correction, etc. While $a>1$, the problem becomes highly
non-linear because of the relativistic effects and it can hardly
be solved.

In full ionized plasmas, there are two dominating nonlinear
effects on the laser propagation: one is the relativistic mass
increase of the electron. When the amplitude of vector potential
$a$ increases, the vibration velocity of electron increases and
changes the mass of electron to $\gamma m$, where the Lorentz
factor $\gamma=\frac 1 {\sqrt{1-v^2/c^2}}$ is a coefficient
determined by the velocity, so non-linear effects appear.

The other important non-linear effect is the laser's diffraction
by electron density waves. When the intense laser propagates,
electrons in the plasma first vibrate parallel to the electric
field and perpendicular to the propagating direction of the laser.
Then the vibrating elections are pushed by the Lorentz force
produced by the magnet field of the laser. Electron density
oscillations parallel to the propagating direction of the laser
form density waves.  Finally, laser is diffracted by the density
waves. As is shown in the following, the density fluctuation is
proportion to $\mathbf{a}\cdot\mathbf{a}$,  its frequency is twice
as the laser frequency and its order is the same as relativistic
effect's order. So the density wave must be considered in the
relativistic region.

In the relativistic region, both non-linear effects cause
corrections of  the refract index of the plasma by modifying the
dispersion relation of the laser. The correction is very important
for the shaping and self-focusing of the intense laser pulse.

There are many  parametric instabilities in plasmas such as Raman
instability, Brillouin instability and $2\omega$ decay, etc. All
these processes will be affected by the relativistic  mass
increase and the density wave effects. Stimulated Brillouin
Scattering(SBS) is a very important phenomenon because it
transfers almost all energy to the scattered light and reaches a
maximum on backward scattering. In the case of backward SBS, it
causes laser's reflection and the energy loss. During the
implosion of ICF, the Brillouin reflectivity can vary from
10\%(for short wave incident laser) to more than 40\% (for long
wave laser)$^{\cite{6}}$. In addition, the Brillouin Instability
relates closely with the filamentation instability.

In recent papers${\cite{8}\cite{9}}$, relativistic mass increase
effects are investigated by introducing a Lorentz factor. However,
because the Lorentz factor varies with the electron's jitter
velocity, it takes on different values for different non-linear
effects and should be calculated separately in different cases.
Besides, the SBS takes place in all the under-dense region of
plasmas, while Stimulated Raman Scatter(SRS) can take place only
in the low density region. Thus the density wave effects should be
considered.

Some authors{\cite{10}\cite{11}} already considered the density
wave effects with the laser pulse propagation without handling the
SBS instability. Besides, the density wave equation should be
altered in the relativistic region (see sec. 2).

The purpose of this paper is to investigate the Brillouin
instability of an intense laser with $a^2\sim 0.1-0.5$. We call
this intensity region as Weak-Relativistic Region. Which indicates
that the nonlinear effects can be processed by the series
expansion with respect to $a^2$. For calculating the Brillouin
effect ,we shall use the quasi-periodic approximation to handle
density wave effects to get non-linear corrections of the laser's
dispersion relation with higher precision in sec. 2. In sec. 3, we
shall investigate the Brillouin instability in the relativistic
region.

\section{The non-linear effects of the intense laser propagating in Plasmas}

Consider the propagation of a linear-polarized laser in a cold
plasma in which the electron's thermal velocity is much smaller
than their jitter speed in the laser field\cite{12}. Suppose the
laser is polarized in the $x$ direction and propagates in the $z$
direction with $a<1$. Assuming the plasma be homogeneous in the
transverse direction, we have, from the conservation of transverse
canonical momentum\cite{2}, $\gamma \beta_x=a_x$ and
$\gamma\doteq\sqrt{1+\mathbf{a}^2}$, where
$\bm{\beta}=\mathbf{v}/c$.

The laser propagation equation in this case has been derived by P.
Sprangle and et al \cite{13} as follows:

\begin{equation}
-\frac {\partial^2 \mathbf{a} }{\partial c^2 t^2}+\frac {\partial ^2 \bf{a}}{\partial z^2}=
k_p^2 \frac n {\gamma n_0}\mathbf{a}
\end{equation}

where $n$ is the density of electrons, $n_0$ is the local average
of $n$, $k_p=\omega_p/c$ and $\omega_p$ is the plasma frequency.

The plasma should satisfy Vlasov equation. In the cold plasma
approximation we ignore the thermal-pressure to get the continuity
equation and the equation of motion from the first two moments of
the Vlasov equation:

\begin{eqnarray}
\label {vlasov1}
\frac {\partial n}{\partial ct}\ +  \frac {\partial (n\beta_z)}{\partial z}=0 \\
\label{vlasov2}
 \frac {\partial (\gamma \beta_z)}{\partial{ct}}+\beta_z \frac {\partial (\gamma \beta_z)}{\partial z}=
\frac {\partial \phi}{\partial z}-\frac 1 {2\gamma}\frac {\partial }{\partial z}{\mathbf{a}^2}
\end{eqnarray}

The above equations cannot be simplified to an ordinary wave
equation as in non-relativistic case due to the presence of
$\gamma$ under derivative operations. The non-linear effects must
be treated separately for high frequency, low frequency and
zero-frequency cases to get the correct dispersion relations.

In the quasi-periodic approximation we assume that the
characteristic time of evolution of the laser pulse shape is much
longer than the laser's period, and $\mathbf{a}$ can be regarded
approximately as a periodic function. Let $\mathbf{a}\doteq a_0
\mathbf{\hat{x}} {\cos(\omega_0 t- k_0 z)}$, where
$\mathbf{\hat{x}}$ is the unit vector in the $x$ direction. Now
all of the coefficients and driving forces in eq. (\ref{vlasov1})
and (\ref{vlasov2}) are periodic functions, so its solutions
should be periodic far from their instability region.

The approximation is similar to the quasi-stationary approximation
used  by P.Sprangle et al $^{\cite{13}}$, but now the phase
velocity of the laser wave in the plasma, $\omega_0/k_0$, does not
need be close to $c$. So this approximation can be used for dense
plasmas.

Introduce the phase variable $\xi=\omega_0 t-k_0 z$ and let
\begin{eqnarray}
\nonumber &n=&n(\xi)\\
\nonumber &\beta_z=&\beta_z(\xi)\\
\nonumber &\phi=&\phi(\xi)
\end{eqnarray}

After simple calculations, (\ref{vlasov2})becomes

\begin{equation}
\frac {\omega_0^2} {k_0 c^2}\frac {n_0} {n}\frac d {d \xi}({\gamma}\frac {n-n_0}n)
=\frac {\partial \phi }{\partial z}-\frac 1 {2\gamma}\frac {\partial }{\partial z}{\mathbf{a}^2}
\end{equation}

Let $n=n_0(1+\psi)$ and use the poisson equation, the linearized
equation in $\psi$ becomes

\begin {equation}
\label{density}
- \frac {\partial^2 (\gamma \psi)} {\partial c^2 t^2}=k_p^2 \psi -\frac \partial {\partial z}
(\frac 1 {2 \gamma} \frac \partial {\partial z}(\mathbf{a}^2))
\end{equation}

Expand the Lorentz factor $\gamma$ and preserve terms lower than
fourth, we get the nonlinear dispersion relation by an iteration
method:

\begin{equation}
\label{dispersion}
 \frac {\omega_0^2}{c^2}-k_0^2-k_p^2(1-\frac 3 8
a_0^2)=\frac 1 2 k_p^2 \psi_0 =\frac {k_p^2} 2 \frac {k_0^2 c^2
a_0^2}{4\omega_0^2(1+\frac 1 4 a_0^2)-\omega_p^2}(1-\frac 1 4
a_0^2)
\end{equation}

Here the relativistic terms appear with effective Lorentz factors
$(1+\frac 14 a_0^2)$ and $(1+\frac 3 8 a_0^2)$ respectively. As we
said before, the efficient Lorentz factors differ for different
modes.The factor $\psi_0\propto k_0^2$ is the frequency-mixing
term. The ratio $\psi_0/k_0^2$ increases while $\omega_0$
approaches $\omega_p$ . When $a_0^2\sim 0.3$ and
$\omega_0^2=2\omega_p^2$,we have $\psi_0\sim 0.04$.

\section{The relativistic correction of Stimulated Brillouin Instability}

Consider a two-dimensional homogenous plasma and ignore the
electron's thermal motion in the direction of laser polarization.
Let $\nabla_\perp=\{\frac {\partial}{\partial x}, \frac
{\partial}{\partial y}\}$, $\nabla^2_\perp=\frac
{\partial^2}{\partial x^2}+\frac {\partial^2}{\partial y^2}$,
$\bm{\beta}=\{v_x/c,v_y/c,v_z/c\}$, then $\nabla^2=\frac
{\partial^2}{\partial z^2}+\nabla_\perp^2$.

Suppose that the incident laser propagates in $z$ direction and is
polarized parallel to $x$ axis, the electron's temperature in the
plasma is $T$, then $kT\ll m_0v_x^2$. The conservation of the
transverse canonical momentum leads to $\gamma v_x/c \approx a_x$.
Now the Lorentz force has the following form $\mathbf{v}\times
(\nabla \times \mathbf{a})\approx \frac 1 {2\gamma} \nabla
(\mathbf{a}\cdot\mathbf{a})$,

So the equation of motion for electrons becomes
\begin{equation}
\frac {\partial}{\partial ct}(\gamma \bm{\beta})+(\bm{\beta} \cdot \nabla)(\gamma \bm{\beta})
=\nabla \phi-\frac 1 {2\gamma} \nabla (\mathbf{a}\cdot\mathbf{a})-\nabla {\frac {P_e}{n_0 m_0 c^2}}
\end{equation}

In stead of $P_e$, we shall introduce the normalized electron
thermal-pressure $p=\frac {P_e}{m_0 c^2}=\frac
{(n_0+\tilde{n})\theta}{m_0 c^2}$, where $\theta=kT$.

In Brillouin Scattering, the plasma acoustic wave is coupled with
the laser wave. Let us consider low frequency fluctuations.
Ignoring the electron's inertia gives

$$\{\nabla \phi\}^{low}=\{\frac 1 {2\gamma}\nabla (\mathbf{a}^2)\}^{low}-\frac {\theta}{m_0 n_0 c^2}\nabla (\tilde{n}^{low})$$
where $\{\}^{low}$ indicates preserving low frequency(acoustic)
terms only.

Substituting this equation into ion's equation of motion, ignoring
ion's thermal pressure($T_i\ll T_e$) and using the relation $Z
n_i=n$ we have:
$$\frac {\partial^2 \tilde{n}^{low}}{\partial c^2t^2}-\frac {Z n_0 m_0 }{M}\left\{[\nabla^2\sqrt{1+\mathbf{a}^2}]^{low}
+\frac {\theta}{m_0 c^2}\nabla^2 (\frac {\tilde{n}^{low}}
{n_0})\right\}=0$$

When $a<1$, $\sqrt{1+\mathbf{a}^2}\approx 1+\frac 1 2
\mathbf{a}^2-\frac 1 8 \mathbf{a}^4$. Let
$\mathbf{a}=\mathbf{a_L}+\tilde{\mathbf{a}}_s$, where
$\mathbf{a}_L$ and $\tilde{\mathbf{a}}_s$ is for the incident
laser and the scattered light respectively.

Now use the relation $|\tilde{\mathbf{a}}_s|\ll|\mathbf{a}_L|$, we
have
\begin{equation}
\label{brillouin1}
\frac {\partial^2 \psi_s}{\partial t^2}-c_s^2
\nabla^2\psi_s= \frac {Z m_0 c^2}{M}\nabla^2(\mathbf{a}_L\cdot
\tilde{\mathbf{a}}_s-\frac 1 2 \mathbf{a_L}^2 \mathbf{a}_L\cdot
\tilde{\mathbf{a}_s})^{low}
\end{equation}
where $\psi_s=\frac {\tilde{n}^{low}} {n_0}$ and $c_s=\sqrt{\frac
{Z\theta}M}$ is the velocity of ion-acoustic wave in the plasma.

Similarly, we can write the density fluctuation in the form of
$\psi=\frac {\tilde{n}}{n_0}=\psi_L+\tilde{\psi_s}$ and set up the
modified propagation equation:
\begin{equation}
\label{brillouin2}
(-\frac {\partial ^2 }{\partial c^2t^2}+\frac
{\partial^2}{\partial
z^2}+\nabla_\perp^2-k_p^2)\tilde{\mathbf{a}}_s= k_p^2[\psi_L
\tilde{\mathbf{a}}_s+\tilde{\psi_s} \mathbf{a}_L]-\frac 3 2 k_p^2
\mathbf{a}_L^2 \tilde{\mathbf{a}}_s
\end{equation}
where $\psi_L$ is the density wave generated  by the laser driving
solely.

The coupled equations (\ref{brillouin1}) and (\ref{brillouin2})
are the fundamental equations for the Stimulated Brillouin
Scattering in the weak relativistic region. There are two
differences between the present equations and the normal Brillouin
equations. First, the factor $\sqrt{1+\mathbf{a}^2}$ have replaced
the factor $\frac 1 2 \mathbf{a}^2$. Second, in the propagation
equation, the density wave term $\psi_L \tilde{\mathbf{a}}_s$ and
the relativistic term $\frac 3 2 \mathbf{a}_L^2
\tilde{\mathbf{a}}_s$ cause a new effect, which is the sideband
mixing.

Assume that an acoustic density-perturbation is formed in the
plasma: $n_s^{low}= n_0 \psi_a \cos (\omega t -k z)$. It is
coupled with the laser field in the plasma and produces the
sideband scattered light waves like
$\tilde{\mathbf{a}}_s=\mathbf{a}_{+}\cos [(\omega_0+\omega) t
-(k_0+k)z]+ \mathbf{a}_{-}\cos [(\omega_0-\omega)t -(k_0-k)z]$. In
the linear region($a_0\ll 1$), the two scattered light waves will
propagate independently and will not affect each other.

Let $\omega_{\pm}=\omega_0 \pm \omega$, $k_{\pm}=k_0\pm k$, in the
relativistic region we discussed, the laser will drive a density
wave $\psi_L$ whose basic frequency is $2\omega_0$. This term will
cause frequency mixing with $\mathbf{a}_+ \cos(\omega_+ t -k_+ z)$
and result in a term whose frequency is $\omega_-$. Similarly, it
will mix with $\mathbf{a}_- \cos(\omega_- t -k_- z)$ and result in
a term with frequency $\omega_+$. Namely, two sideband light waves
will be coupled to the basic mode. The term $\mathbf{a}_L^2$ in
$\frac 3 2 \mathbf{a}_L^2 \tilde{\mathbf{a}}_s$ will act in the
same way.

Because of the sideband mixing, $\mathbf{a}_+$ and $\mathbf{a}_-$
must be handled together to get cooperative dispersion relations
of SBS.

We consider the simplest sideband mixing effects in order to get
rid of the unnecessary complexity. Namely, we consider the mixing
effects caused by $\psi_L \propto \cos(2\omega_0 t-2 k_0z)$ and
$\mathbf{a_L^2}$ only. Assuming the incident laser propagates
parallel to the $z$ axis and the scattered light propagate in the
$y$-$z$ plane, the incident wave vector is $\{0,0,k_0\}$, the two
scattered sideband wave vectors are $\{0,k_\perp,k_0+k\}$ and
$\{0,-k_\perp,k_0-k\}$, we can write the incident laser in the
form of $\mathbf{a}_L=a_0 \hat{\mathbf{x}}\cos(\omega_0 t-k_0 z)$
, write the scattered light as
$\tilde{\mathbf{a}_s}=\hat{\mathbf{x}}\{a_{+}\cos
[(\omega_0+\omega) t -(k_0+k) z-k_\perp y]+ a_{-}\cos
[(\omega_0-\omega) t -(k_0-k)z+k_\perp y]\}$
 and write the acoustic density fluctuation as $\tilde{\psi_s}=\psi_a \cos(\omega t-k z-k_\perp
y)$. The sideband terms in the rhs of the propagation equation
$$(-\frac {\partial ^2 }{\partial c^2t^2}+\frac {\partial^2}{\partial z^2}+\nabla_\perp^2-k_p^2)\tilde{\mathbf{a}}_s=
k_p^2[\psi_L \tilde{\mathbf{a}}_s+\tilde{\psi_s}
\mathbf{a}_L]-\frac 3 2 k_p^2 \mathbf{a}_L^2
\tilde{\mathbf{a}}_s$$ are:

Sideband terms in $k_p^2\psi_L\tilde{\mathbf{a}_s}$ are
$$k_p^2\psi_L\tilde{\mathbf{a}_s}\longrightarrow \frac 12 k_p^2\psi_0\{a_-\cos(\omega_+t-k_+z-k_\perp y)
+a_+\cos(\omega_-t-k_-z+k_\perp y)\}$$

Sideband terms in $k_p^2\tilde{\psi_s}\mathbf{a}_L$ are
$$k_p^2\tilde{\psi_s}\mathbf{a}_L\longrightarrow\frac {k_p^2}{2}\psi_a a_0 \{\cos(\omega_+t-k_+z-k_\perp y)
+\cos(\omega_-t-k_-z+k_\perp y)\}$$

Finally, sideband terms in $-\frac 3 2
k_p^2\mathbf{a}_L^2\tilde{\mathbf{a}_s}$ are
\begin{eqnarray}
\nonumber -\frac 3 2
k_p^2\mathbf{a}_L^2\tilde{\mathbf{a}_s}\longrightarrow-\frac 34
k_p^2 a_0^2[a_+\cos(\omega_+t-k_+z-k_\perp y)
+a_-\cos(\omega_-t-k_-z+k_\perp y)]\\
\nonumber-\frac 3 8 k_p^2 a_0^2a_+\cos(\omega_-t-k_-z+k_\perp y)-\frac 38 k_p^2 a_0^2 a_-\cos(\omega_+t-k_+z-k_\perp y)
\end{eqnarray}

Matching all the sideband terms, we get:
\begin{eqnarray}
\bigg{\{}
\begin{array}{cc}
D_+a_+=\frac 12k_p^2\psi_0a_-+\frac 12 k_p^2\psi_a a_0-\frac 34k_p^2a_0^2a_+-\frac 38 k_p^2a_0^2a_-\\
D_-a_-=\frac 12k_p^2\psi_0a_++\frac 12 k_p^2\psi_a a_0-\frac 34k_p^2a_0^2a_--\frac 38 k_p^2a_0^2a_+
\end{array}
\end{eqnarray}
where $D_\pm =\frac {\omega_\pm^2} {c^2}-k_\pm^2-k_\perp^2-k_p^2$.

$a_+$ and $a_-$ can be solved as
\begin{eqnarray}
\Bigg{\{}
\begin{array}{ccc}
\nonumber a_+=\frac {(D_-+\frac 3 4 k_p^2 a_0^2)+(\frac 1 2 k_p^2 \psi_0 -\frac 3 8 k_p^2 a_0^2)}
{(D_++\frac 3 4 k_p^2 a_0^2)(D_-+\frac 3 4 k_p^2 a_0^2)-(\frac 1 2k_p^2\psi_0-\frac 3 8k_p^2a_0^2)^2}
\frac 12k_p^2\psi_a a_0\\
\nonumber a_-=\frac {(D_++\frac 3 4 k_p^2 a_0^2)+(\frac 1 2 k_p^2 \psi_0 -\frac 3 8 k_p^2 a_0^2)}
{(D_++\frac 3 4 k_p^2 a_0^2)(D_-+\frac 3 4 k_p^2 a_0^2)-(\frac 1 2k_p^2\psi_0-\frac 3 8k_p^2a_0^2)^2}
\frac 12k_p^2\psi_a a_0
\end{array}
\end{eqnarray}
The driving density wave equation is
\begin{eqnarray}
\nonumber \frac {\partial^2 \tilde{\psi_s}}{\partial t^2}-c_s^2 \nabla^2 \tilde{\psi_s}=
\frac {Z m_0c^2}{M}\nabla^2\left(\mathbf{a}_L\cdot\tilde{\mathbf{a}_s}-\frac 1 2 \mathbf{a}_L^2 \mathbf{a}_L\cdot\tilde{\mathbf{a}_s}\right)^{low}\\
=-\frac{Zm_0 c^2}{M}\frac 1 2 a_0(a_++a_-)(1-\frac 3 8a_0^2)(k^2+k_\perp^2)\cos(\omega t-kz-k_\perp y)
\end{eqnarray}
Substituting the above expression for $a_\pm$ into it, we get the
Stimulated Brillouin Scattering's dispersion relation:
\begin{eqnarray}
\label {bri_dispersion}
\nonumber\omega^2-c_s^2(k^2+k_\perp^2)=\frac {Z m_0 c^2}{M}\frac 1
4 k_p^2(k^2+k_\perp^2)
(1-\frac 3 8 a_0^2)a_0^2\\
\times \frac {(D_++D_-)+k_p^2(\psi_0+\frac 34a_0^2)}
{(D_++\frac 34 k_p^2a_0^2)(D_-+\frac 3 4 k_p^2 a_0^2)-[\frac 12 k_p^2(\psi_0-\frac 34 a_0^2)]^2}
\end{eqnarray}

This equation can be solved numerically. Let us consider the
backward Brillouin scatter$\cite{1}$, when $k\simeq 2k_0$ and
$k_\perp=0$. Let $\omega=\Omega \kappa_p c$,$k=\kappa k_p$,and
$\Omega_0=\omega_0/{\omega_p}$, $\kappa_0=k_0/{k_p}$ ,we normalize
the equation $(\ref{bri_dispersion})$ to
\begin{eqnarray}
\nonumber &(\Omega^2-\kappa^2)&=\frac {Z m_0}{M} \frac 1 2\kappa^2(1-\frac 3 8 a_0^2)a_0^2\\
& &\times \frac {\Omega^2 \frac
{c_s^2}{c^2}-\kappa^2+\psi_0}{[\Omega^2\frac
{c_s^2}{c^2}-\kappa^2+\frac 12 (\psi_0+\frac 34 a_0^2)]^2-4
[\Omega \Omega_0\frac {c_s^2} {c^2}-\kappa \kappa_0]-\frac 1
4(\psi_0-\frac 3 4 a_0^2)^2}
\end{eqnarray}

We adopt the typical parameters $\frac{c_s}{c}=10^{-3}$,$\frac M
m=1800$ ,the intensity of laser is set to $a_0^2=0.32$ ,the plasma
density is set to $\omega_0^2=4\omega_p^2$ and
$\omega_0^2=2\omega_p^2$ respectively. The results are plotted on
figure 1 and figure 2. In the weak relativistic region,the ion
acoustic waves become the quasi-mode whose frequency is determined
by the laser intensity. Correspondingly, the relative value of
growing rate and the laser frequency will far exceed the ratio
$c_s/c$ and approach to $\frac 1 {40}$ in the region of the
parameter we selected. It means that the Brillouin instability
will have developed during about several tens to a few hundreds
periods of the pump laser(about $10^{-13}-10^{-14}$ seconds ).

In the region we analyzed,the pure growing modes with
$Re(\omega)=0$ appear at large $k$. First, The peak value of the
growing rate locates near $k=2k_p$ and has a departure due to the
relativistic effects. Then, the pure growing modes appear at large
$k$ when the laser intensity and the plasma density increase. In
the weak relativistic region, the pure growing modes form a
plateau region. The pure growing mode at large $k$ will affect on
some nonlinear effects like turbulence developing.
\begin{figure}
\includegraphics{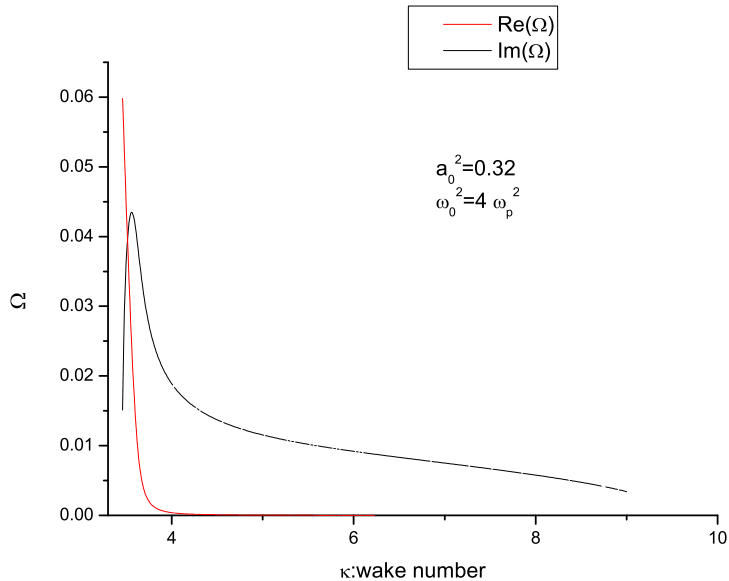}
\caption{the growing rate when $n=\frac 1 4 n_0$}.
\end{figure}

\begin{figure}
\includegraphics{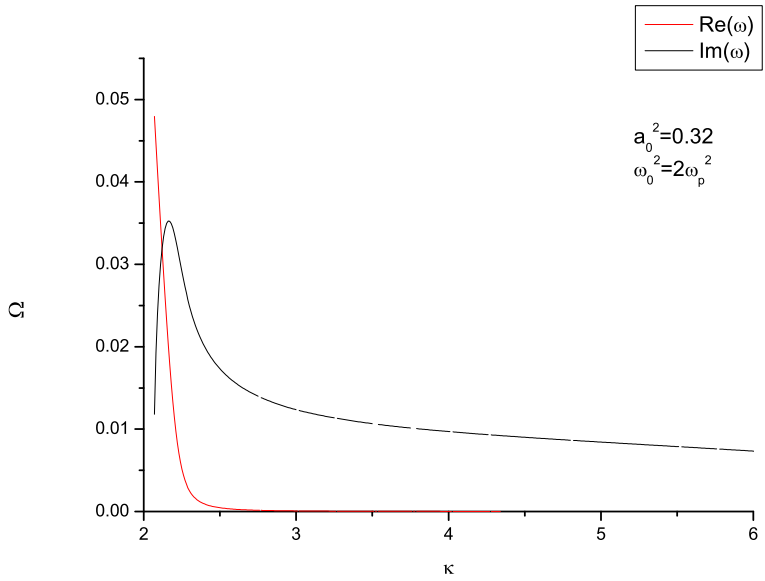}
\caption{the growing rate when $n=\frac 1 2 n_0$}.
\end{figure}

Ignore the relativistic terms and the frequency-mixing terms, the
dispersion relation $(\ref{bri_dispersion})$ returns to the
non-relativistic four-wave dispersion relation of Brillouin
scatter. We can solve the non-relativistic dispersion relation and
compare the result with the relativistic result. All the results
are plotted in figure 3. In which we can find the effects of
relativity and frequency-mixing are: (1)move the peak value;
(2)reduce the growing rate of instability. The reduction is larger
in the plateau region of pure growing modes. So the Brillouin
reflection rate will be reduced by the effects.

\begin{figure}
\includegraphics{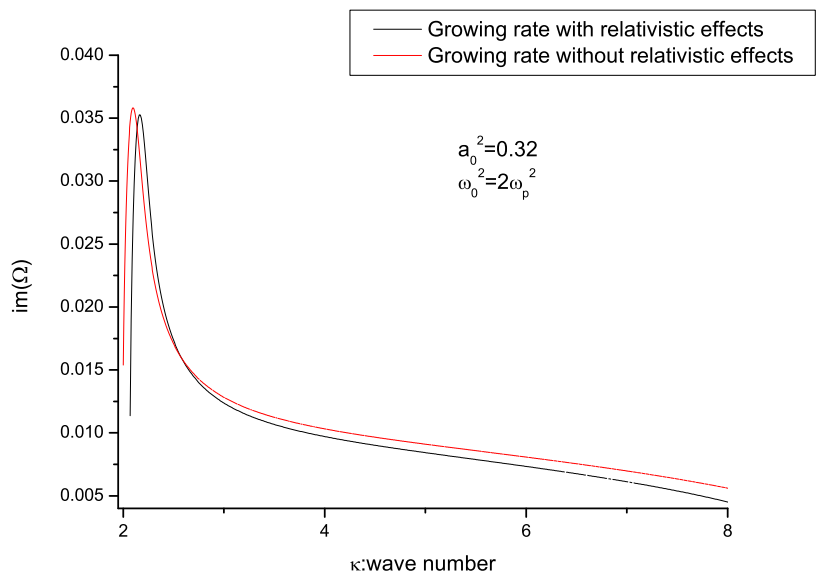}

\caption{the comparison of relativistic and non-relativistic
growing rates.}
\end{figure}

\section{Conclusion}

The present paper applied the quasi-periodic approximate method
for the stable laser propagation in weak relativistic plasmas. The
Stimulated Brillouin Scatter in relativistic region is
investigated in detail. The growing rates are re-calculated
numerically and the sideband-mixing effects are considered. The
Brillouin instability is reduced by these effects. For the plateau
region of pure growing modes, the reduction is more evidence.

\section*{reference}

\end{document}